\documentclass[sigconf]{acmart}

\PassOptionsToPackage{table}{xcolor}

\AtBeginDocument{%
  }

\usepackage{graphicx}
\usepackage{subcaption}
\usepackage{multirow}
\usepackage[ruled,vlined]{algorithm2e}
\usepackage{adjustbox}

\usepackage{tabularx}
\usepackage[table]{xcolor} 
\usepackage{enumitem}

\usepackage{booktabs}
\usepackage{array}
\usepackage{colortbl}
\usepackage{arydshln}

\newcolumntype{L}{>{\RaggedRight\arraybackslash}X} 

\begin{document}

\title{Multi Interests for Joint Search-Recommendation Modeling}


\copyrightyear{2027}
\acmYear{2027}
\setcopyright{cc}
\setcctype{by}

\acmConference[WSDM 2027]
{Proceedings of the Twentieth ACM International Conference
on Web Search and Data Mining}
{February 15--19, 2027}
{Hong Kong, Hong Kong}

\acmBooktitle{Proceedings of the Twentieth ACM
International Conference on Web Search and Data Mining
(WSDM 2027), February 15--19, 2027, Hong Kong, Hong Kong}

\acmISBN{979-8-4007-3117-4/2027/02}
\acmDOI{10.1145/3862322.3863806}

\settopmatter{printacmref=true}

\ccsdesc[500]{Information systems~Personalization}
\ccsdesc[500]{Information systems~Recommender systems}
\ccsdesc[300]{Information systems~Learning to rank}


\author{Xiangchen Pan}
\orcid{0009-0008-4641-5531}
\affiliation{%
  \institution{Huazhong University of Science and Technology}
  \city{Wuhan}
  \country{China}}
\email{m202474126@hust.edu.cn}

\author{Wei Wei}
\orcid{0000-0003-4488-0102}
\authornote{Corresponding author.}
\affiliation{%
  \institution{Huazhong University of Science and Technology}
  \city{Wuhan}
  \country{China}
}
\email{weiw@hust.edu.cn}

\begin{abstract}
Search and recommendation are crucial for understanding user preferences. More and more studies are attempting to jointly model search behavior and recommendation behavior, by integrating user active search and passive recommendation behavior data to better mine user preferences. However, although existing cross-domain unified modeling frameworks can effectively compensate for the differences in behavior between domains, they overlook the expression of interests in different scenarios under mixed sequences. In this study, we propose a multi-interest-based mixed sequential modeling framework MIJSR, which performs multi-interest mining and adaptive integration on search recommendation mixed sequences from both structural and semantic perspectives. Specifically, our model can be roughly divided into three modules: cross-domain behavior fusion, multi-interest mining, and multi-task prediction. Firstly, we align the representations of query and item through contrastive learning training. Then, we extract the multi interests of the mixed behavior sequence from both structural and semantic perspectives. Structurally, we extract search interests, recommendation interests, and cross interests through subsequence partitioning and mask settings; In terms of semantics, we use the semantic information of queries for clustering and perform semantic segmentation on mixed sequences to construct semantic multi interests. Finally, the adaptive fusion of multiple interests is combined with other side information to use a progressive layered extraction model for multi-task prediction. Extensive experiments on two open-source datasets have shown that our model can further enhance its accuracy in search and recommendation by extracting users' multi interests at a fine-grained level. Codes are available at https://github.com/pxcstart/MIJSR.
\end{abstract}

\keywords{Personalized Search, Sequential Recommendation, Joint Search and Recommendation Model}


\maketitle

\section{Introduction}

Search engines and recommendation systems (S\&R) have become important tools for users to obtain information~\cite{bennett2012modeling, cheng2016wide, ge2018personalizing, lu2019psgan, wu2019npa}. Users can initiate a search in the search box or interact with the displayed items on the recommendation interface. Traditionally, search and recommendation have been regarded as two independent research scenarios. In recent years, many social or e-commerce platforms have tended to jointly model users' search behavior and recommendation behavior~\cite{shi2024unisar, shi2025gensar, xie2024unifiedssr, zamani2020learning}, which can reduce data sparsity in a single scenario and provide a more comprehensive understanding of user intentions.

Early joint modeling of search and recommendation predominantly followed a representation-first, fusion-later paradigm: separate encoders were used to extract user interests from search and recommendation behaviors, respectively (e.g., NRHUB~\cite{wu2019neural}, SESRec~\cite{si2023search}). However, this design struggles to capture the intrinsic correlations between search and recommendation. USER~\cite{yao2021user} introduced a unified paradigm based on hybrid sequential modeling, which chronologically interleaves active search and passive recommendation behaviors to better exploit their similarities, and performs long- and short-term interest modeling over the hybrid sequence to uncover user preferences. UniSAR~\cite{shi2024unisar} further emphasizes fine-grained transitions between search and recommendation behaviors, and strengthens cross-domain interactions via cross-attention.


Although these methods can largely bridge behavioral differences across domains, they overlook multi-interest expression under a hybrid-sequence view. For example, a user with long-term preferences for digital devices and designer toys, due to an upcoming trip to Japan, begins frequently searching for travel-related queries such as “Tokyo hotels” and “Japan visa" while clicking items in the recommendation stream such as suitcases, cameras, and noise-canceling headphones. If search and recommendation behaviors are directly modeled as a unified mixed sequence, the model tends to naively entangle interests such as travel, digital, and designer toys, causing short-term travel intent to be overshadowed by long-term historical interests. By contrast, extracting multi-interest representations from multiple perspectives, such as search interests, recommendation interests, and search–recommendation cross-interests, can  model users' current travel intentions, long-term digital preferences, and "technology consumption needs in travel scenarios" separately. This helps mitigate interest coupling caused by heterogeneous behavior mixing, and enables a more accurate characterization of dynamic needs and personalized preferences.

In addition, in previous work, mixed sequence modeling 	
was primarily based on ID features for feature interaction and interest expression. The utilization of query features only stays at the level of mean pooling for word embeddings~\cite{xie2024unifiedssr, shi2024unisar}, and search behavior often directly integrates query embeddings with corresponding click item embeddings, which weakens the original semantic expression of the query. In fact, the semantic information of user search queries can most directly reflect user preferences. For an example, based on the user's search history, "mechanical keyboard" and "wireless mouse" correspond to computer peripheral interests, while "running shoes" and "sports watch" correspond to sports interests. Compared to item IDs, queries naturally provide clearer interest boundaries, so multi-interest extraction based on query semantic clustering can more effectively characterize users' multiple preferences. To this end, we perform semantic clustering on all queries in the dataset in advance. After aligning the features of the queries and items, we perform semantic partitioning on the mixed sequence, dividing it into multiple interests from a semantic perspective.

In summary, in order to better decouple the interests of S\&R mixed sequences, we propose a multi-interest-based joint modeling framework for search and recommendation, MIJSR. The model mainly consists of three parts: cross-domain behavior fusion, multi-interest mining, and multi-task prediction. Considering that search scenarios provide additional explicit search queries compared to recommendation scenarios, we designed a contrastive learning task of query and item for feature alignment. Next, we extract multi interests from mixed sequences from both structural and semantic perspectives. On the structural side, we extract search interests, recommendation interests, and cross interests by dividing subsequences and setting masks; On the semantic side, we use the semantic information of queries for clustering and extract semantic subsequences for each category from the mixed sequence to construct semantic multi interests; Finally, we adaptively integrate interests and input them along with other side information into a progressive layered extraction (PLE) model for multi-task prediction.

The main contributions of the paper are as follows:
\begin{itemize}[leftmargin=*] 
    \item We propose a multi-interest-based search-recommendation joint modeling framework MIJSR, and we construct users' multi-interests from both structural and semantic perspectives, comprehensively characterizing users' personalized preferences.
    \item On the structural side, we focus on search interests, recommendation interests, and cross interests; on the semantic side, we use the semantic information of queries for semantic clustering, perform semantic segmentation on mixed sequences, and construct semantic multi-interest.
    \item We conducted experiments on two publicly available datasets, and MIJSR achieved even better performance than state-of-the-art models in both search and recommendation tasks.
\end{itemize}


\section{Related Work}

\subsection{Sequential Recommendation}
Sequential recommendation is based on user history sequences generated in chronological order, modeling the dynamic changes in user interests and predicting the user's next interaction. Early sequential modeling used Markov chains~\cite{rendle2010factorizing}, but with the advancement of deep learning, GRU4Rec~\cite{hidasi2015session} and Caser~\cite{tang2018personalized} applied RNN~\cite{li2017neural} and CNN~\cite{lecun1998gradient} for sequence modeling. Later, with the widespread application of Transformer~\cite{devlin2019bert} in fields such as NLP, SASRec~\cite{kang2018self} and Bert4Rec~\cite{sun2019bert4rec} validated its enormous potential in sequential recommendation. In the industrial sector, the intentions of users with long sequences may be diverse. DIN~\cite{zhou2018deep} introduces an attention mechanism to weight and aggregate historical behavior based on candidate items, in order to model multiple user interests; KA-MemNN~\cite{zhu2020sequential} uses item categories as intent proxies with memory networks; DSIN~\cite{feng2019deep} divides the sequence into sessions to characterize the changes in users' interests in different sessions; The SDM~\cite{lv2019sdm} model has been further extended on the basis of DSIN, dividing user history sequences into long-term and short-term sessions for extracting long-term and short-term interests respectively. In this work, we perform cross-domain sequence modeling on search-recommendation mixed sequences. We not only explore users' multiple interests in sequence structure, but also consider dividing semantic multi interests based on query semantics.

\subsection{Personalized Search} 
Personalized search aims to retrieve items that are not only relevant to the current query but also consistent with users' individual preferences. Early studies mainly focused on query-item relevance matching, such as QEM~\cite{ai2019zero} and DREM~\cite{ai2019explainable}, which estimate user interests solely based on semantic similarity between queries and items. To better capture personalized intent, subsequent works incorporated user profiles and historical search behaviors into the retrieval process. HEM~\cite{ai2017learning} introduced user embeddings to model long-term preferences through review information, while AEM employed an attention mechanism to aggregate historical interactions according to the current query. Building upon AEM, ZAM~\cite{ai2019zero} introduced a zero-attention vector to adaptively control the degree of personalization. More recently, TEM~\cite{bi2020transformer} replaced the attention layer with a Transformer encoder to capture sequential dependencies in users' search histories. In addition, several studies have explored multi-interest modeling in personalized search. For example, CAMI~\cite{liu2022category} leverages knowledge graph embeddings to disentangle users' multiple interests, while GraphSRRL~\cite{liu2020structural} exploits structural patterns in user-query-item graphs to enhance interest representation. Although these methods have achieved promising results, they primarily rely on search behaviors and overlook valuable preference signals contained in recommendation interactions. Consequently, their ability to model user interests is limited in sparse-search scenarios. Different from existing personalized search methods, our work incorporates recommendation behaviors into the search process and performs unified multi-interest modeling over search and recommendation interactions, enabling a more comprehensive understanding of user preferences.



\subsection{Joint Search and Recommendation}
In recent years, joint modeling of search and recommendation scenarios has gradually become a major trend for many commercial platforms. Early work mainly adopted a strategy of representation before fusion, that is, modeling the query sequence and item sequence separately, and then using search data as supplementary information to improve recommendation performance. For example, NRHUB~\cite{wu2019neural} proposed an attention multi-view learning framework that uses a hierarchical attention masker to learn a unified user representation from heterogeneous behaviors; Query-SeqRec~\cite{he2022query} builds heterogeneous sequences based on queries to predict the next interaction; IV4Rec~\cite{si2022model} uses causal learning to treat search queries as instrumental variables for decomposing embeddings and mitigating bias; SESRec~\cite{si2023search} utilizes contrastive learning to distinguish between similar and different representations between search and recommendation. However, these methods overlook the intrinsic connection between search and recommendation tasks, mainly focusing on utilizing search data to improve recommendation performance, and the search performance of the model has not significantly improved.

Recently, more and more work has shifted towards unified modeling of search and recommendation data, aiming to promote joint enhancement between search and recommendation. USER~\cite{yao2021user} integrates the user's behavior during the search and recommendation process into a heterogeneous behavior sequence, which is then encoded using a hierarchical encoder; UniSAR~\cite{shi2024unisar} models user transition behavior between search and recommendation through three steps of extraction, alignment, and fusion, utilizing cross attention to extract transitions between different behaviors. In this work, we integrate search and recommendation behaviors based on previous work experience, construct a mixed sequence for joint modeling, and extract users' multiple interests in cross-domain scenarios to achieve personalized search and recommendation.

\section{Preliminaries}
$U$, $I$, and $Q$ represent the sets of users, items, and queries, respectively. Each user $u \in U$ has an interaction history $S_u$ arranged in chronological order, which records all their past search and recommendation behaviors. We denote all search behaviors as $S_u^s$ and all recommendation behaviors as $S_u^r$. $S_u=\{(x_1,b_1),(x_2,b_2),...,(x_N,b_N)\}$. Among them, $N$ represents the length of the sequence, and $b_t$ represents the type ($b_t=1$ represents recommendation behavior, and $b_t=0$ represents search behavior). The following is the definition of the behavioral expression for $x_t$:

\[
x_t =
\begin{cases}
i_t, & \text{if } b_t = 1 \text{ (recommendation)}, \\
\langle q_t, c_{q_t} \rangle, & \text{if } b_t = 0 \text{ (search)} ,
\end{cases}
\]

Where $i_t$ is the item corresponding to the recommended behavior, $q_t$ is the query corresponding to the search behavior, $c_ {q_t}=\{i_1, i_2,..., i_{N_{q_t}}\}$ represents the list of items clicked by the user after searching $q_t$.

The interactive data used in the joint modeling of search and recommendation consists of two parts: search and recommendation, $D=D_R \cup D_S$, where $D_R=\{u, i_{N+1}, S_u, y_{N+1}^R\}$ represents the recommendation dataset, and $D_S=\{u, i_{N+1}, q, S_u, y_{N+1}^S\}$ represents the search dataset. $i_{N+1}$ represents candidate items, and $y_{N+1}$ represents the label of $i_{N+1}$ (0 represents negative samples, 1 represents positive samples). Compared to recommendation scenarios, search scenarios have an additional current search query $q$ when predicting the next item. The training goal is to find an optimization function $f_{\Theta}$ that predicts $\hat{y}^{R}_{u,i_{N+1}}$ and $\hat{y}^{S}_{u,i_{N+1},q}$ based on the interaction history $S_u$. The formal definition of $f_{\Theta}$ is as follows:

\[
\left[
\hat{y}^{R}_{u,i_{N+1}},
\hat{y}^{S}_{u,i_{N+1},q}
\right]
=
f_{\Theta}(u, i_{N+1}, q, S_u),
\]

\section{Methodology}
\begin{figure}[t]
    \centering
    \includegraphics[width=1\linewidth]{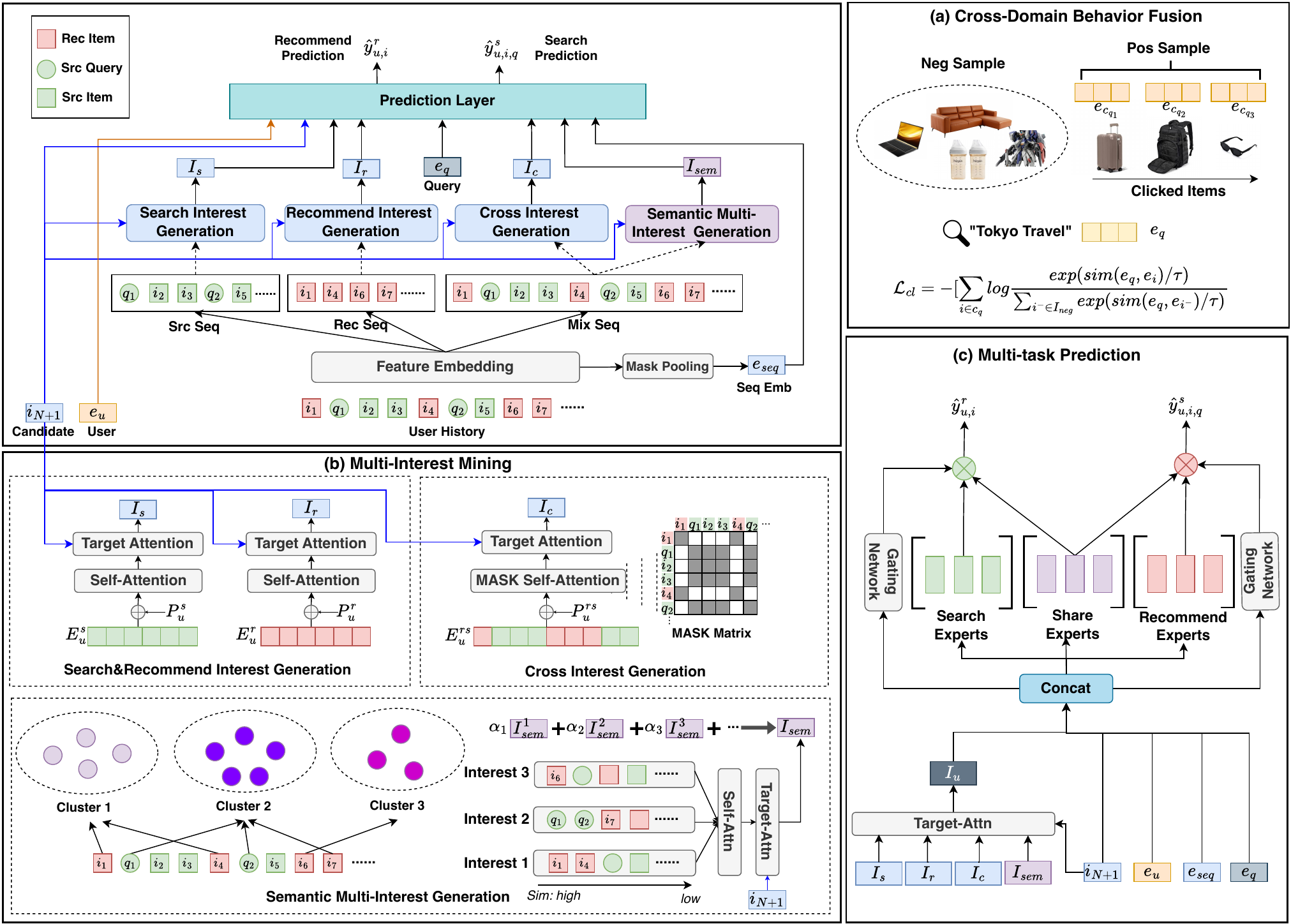}
    \caption{The overview framework of MIJSR. It mainly consists of three modules. Firstly, align the query and item representations through contrastive learning to enhance the behavior representation; Then, for the mixed sequence of search and recommendation, we extract search interests, recommendation interests, and cross interests in structure, and semantically, we divide the mixed sequence into semantic subsequences based on query semantic clustering for extracting semantic multi interests; Finally, we use a multi-expert network to perform multi-task prediction based on user interests and other side information.
}
    \label{overview}
\end{figure}

Fig. \ref{overview} shows the workflow of the MIJSR framework. The framework consists of three parts: cross-domain behavior fusion, multi-interest mining, and multi-task prediction. Firstly, perform contrastive learning on queries and items to enhance behavioral representation; Next, extract users' multiple interests from both structural and semantic aspects of the search and recommendation mixed sequence; Finally, after adaptive fusion of multiple interests, we use a progressive layered extraction model for multi-task prediction. We will give a detailed introduction about each module in the following subsections.

\subsection{Cross-Domain Behavior Fusion}
In terms of embedding representation, users, items, and queries each correspond to a learnable embedding table $E_u \in R^{|U|\times d}$, $E_i\in R^{|I|\times d}$, $E_q\in R^{|Q|\times d}$. Each query $q=[w_1, w_2,...,w_{|q|}]$ is composed of multiple words, and we also maintain an additional word embedding table $E_w \in R^{|W|\times d}$. The query representation $e_q$ is mainly obtained by average pooling the word list $[w_1, w_2,...,w_{|q|}]$. The formal definition is: $e_q=Mean(e_{w_1}, e_{w_2},..., e_{w_{|q|}})$.

For the user interaction history $S_u$, each recommendation behavior is represented as a corresponding item embedding. When constructing representations for search behaviors, the clicked-item embeddings are first mean-pooled, and the pooled vector is then added to the query embedding. The following provides a formal definition for constructing behavioral representations:

\begin{equation}
\begin{aligned}
e_{x_t} &= \left\{
\begin{array}{ll}
e_{i_t}, & \text{if } b_t=1 (Recommend) \\
e_{q_t}+Mean(c_{q_t}), & \text{if } b_t=0 (Search)
\end{array}
\right. 
\end{aligned}
\end{equation}

Where $Mean(c_ {q_t})=Mean(e_{i_1}, e_{i_2},..., e_{i_{N_{q_t}}})$, used to represent the average embedding of the clicked items corresponding to the search behavior.

Considering that search behavior, compared to recommendation behavior, also has search queries, and queries and items do not share embedding tables when constructing features, in order to better utilize search queries in unified modeling, it is first necessary to align the features of queries and items. Here, we align the feature distributions of query and item through contrastive learning. For the query, we will take the corresponding clicked item $c_q$ as the positive sample, and randomly sample the remaining items to construct the negative sample; the same operation applies to the item. The following is the calculation formula for contrastive learning:

\begin{equation}
\begin{aligned}
\mathcal{L}_{cl} = -[\sum_{i \in c_q}log\frac{exp(sim(e_q,e_i)/\tau)}{\sum_{i^- \in I_{neg}}exp(sim(e_q, e_{i^-})/\tau)}\\
\quad +\sum_{i \in c_q}log\frac{exp(sim(e_q,e_i)/\tau)}{\sum_{q^- \in Q_{neg}}exp(sim(e_{q^-}, e_i)/\tau)}]
\end{aligned}
\end{equation}

Where $\tau$ is the learnable temperature coefficient, and $sim(a, b)=aW_b^T$ is used to calculate the similarity between $a$ and $b$.

\subsection{Multi-Interest Mining}
We conduct multi-interest mining on the mixed sequence obtained by combining search behavior and recommendation behavior from both structural and semantic perspectives. Structurally, we mainly focus on user interests within and between domains by dividing subsequences and setting masks based on behavior types; In terms of semantics, we use the semantic information of search queries for semantic clustering, dividing the mixed sequence into multiple semantic subsequences and constructing semantic multi-interests.

\subsubsection{Multi Interests in Structure}
Considering that search is an active behavior and recommendation is a passive behavior, the former reflects more of the user's immediate interests when entering search terms, while recommendation is generally a push mechanism summarized by the system based on the user's interaction history, which often hits the user's long-term interests. In addition, cross-domain behavior transformation can also reflect users' cross interests, often seen as the evolution of users from clear needs to potential preferences. We will extract recommendation interests and search interests based on recommendation sequences and search sequences, respectively, and extract cross interests by setting attention masks on the mixed sequences. Below, we will provide a detailed introduction to the extraction methods for each interest.

Firstly, we capture the search interest $I_s$ and recommendation interest $I_r$ based on the search sequence $S_u^s$ and recommendation sequence $S_u^r$ of user $u$, respectively, to capture the connections within the sequence. Where $S_u^s=[x_1, x_2,..., x_{N_s}]$, $S_u^r=[x_1, x_2,..., x_{N_r}]$. The original features of the sequence are $E_u^s=[e_1,e_2,...,e_{N_s}]$, $E_u^r=[e_1,e_2,...,e_{N_r}]$. To capture the sequential relationship, we introduced positional encoding $P_u^r$ and $P_u^s$, and the final sequence input is:

\begin{equation}
\begin{aligned}
\hat{E}_u^s=E_u^s + P_u^s, \quad \hat{E}_u^r=E_u^r + P_u^r
\end{aligned}
\end{equation}

Next, we will enhance the connection between behaviors within the domain of the two sequences through a multi-head self-attention mechanism. Taking the search sequence as an example, we will use $\hat{E}_u^s$ as input, the expression ability of the sequence is enhanced through Multi-Head Self-Attention (MHA) and Feed Forward Layer (FFN), respectively. Where $Q=K=V=\hat{E}_u^s$. The formal definition is given below:

\begin{equation}
\begin{aligned}
H_u^s=FFN_s(MHA_s(\hat{E}_u^s,\hat{E}_u^s,\hat{E}_u^s)), \quad H_u^r=FFN_r(MHA_r(\hat{E}_u^r,\hat{E}_u^r,\hat{E}_u^r))
\end{aligned}
\end{equation}

Afterwards, we embed the features of the candidate item $i_{N+1}$ into $e_{i_{N+1}}$ and capture the search interest $I_s$ and recommendation interest $I_r$ through target attention. The formal definition is:

\begin{equation}
\begin{aligned}
w_s &= softmax(H_u^sW_se_{i_{N+1}}), \quad I_s = w_sH_u^s \\
w_r &= softmax(H_u^rW_re_{i_{N+1}}), \quad I_r = w_rH_u^r
\end{aligned}
\end{equation}

Where $W_s \in R^{d\times d}$ and $W_r\in R^{d\times d}$ are learnable parameters. $w_s$ and $w_r$ are attention weights, which reflect the similarity between each item in the sequence and the candidate item $i_{N+1}$. The more similar, the greater the weight.

Next, we extract cross interests from mixed sequences by setting masks. We first add positional encoding to the user's complete interaction history $S_u$ to obtain an enhanced sequence input $\hat{E}_u$:

\begin{equation}
\begin{aligned}
\hat{E}_u=E_u + P_u
\end{aligned}
\end{equation}

In order to capture the inter-domain behavioral connections between search and recommendation, we set up an attention mask matrix $M \in \{0,1\}^{N \times N}$, which shields attention operations for intra-domain behaviors and only focuses on the mutual conversion between search and recommendation behaviors. The value of $M$ is as follows:

\begin{equation}
\begin{aligned}
M_{ij} &= \left\{
\begin{array}{ll}
0, & \text{if } b_i=b_j \\
1, & \text{if } b_i\neq b_j
\end{array}
\right. 
\end{aligned}
\end{equation}

We use the constructed attention mask matrix $M$ for heterogeneous mixed sequence $\hat{E}_u$ and perform masked self-attention operations to enhance interaction between domains:

\begin{equation}
\begin{aligned}
&H_u^{rs}=FFN_{rs}(MHA_{rs}(\hat{E}_u,\hat{E}_u,\hat{E}_u, M)) \\
&MHA_{rs}(\hat{E}_u,\hat{E}_u,\hat{E}_u, M)=softmax(\frac{\hat{E}_u\hat{E}_u^T}{\sqrt{d/h}} \odot M)\hat{E}_u
\end{aligned}
\end{equation}

Similarly, we perform target attention calculation between $H_u^{rs}$ and $e_{i_{N+1}}$ for obtaining the cross interest $I_{cross}$.

\begin{equation}
\begin{aligned}
w_{rs} = softmax(H_u^{rs}W_{rs}e_{i_{N+1}}), \quad I_{cross} = w_{rs}H_u^{rs} 
\end{aligned}
\end{equation}

Where $W_{rs}$ is the learnable parameter, and $w_{rs}$ is the attention score.

\subsubsection{Multi Interests in Semantic}
Since queries are the most direct reflection of user preferences, we first perform semantic clustering on all search queries. We use K-Means as the clustering method, with a total of $k$ cluster centers, and the feature representation of each cluster center is $CQ=[\hat{c}_{q_1},\hat{c}_{q_2},...,\hat{c}_{q_k}]$. Next, we will divide the mixed sequence into semantic subsequences. Due to the lack of textual information on the item side of the dataset and the fact that the id feature embedding and the query word embedding are not in the same feature space, we have designed a multi-layer perceptron $g_\theta$ to map the item features to the semantic space of the query.

In order to train $g_\theta$ for semantic alignment, we designed a sequential level alignment task. We extracted the corresponding query sequences $S_u^q=[q_1,q_2,...,q_n]$ and item sequences $S_u^{c}=[c_{q_1},...,c_{q_n}]$ from the user's search history $S_u^s=[(q_1,c_{q_1}),...,(q_n,c_{q_n})]$. Where the semantic representation of the query sequence $S_u^q$ is $E_u^q=[e_{q_1},e_{q_2},...,e_{q_n}]$; the semantic representation of the item sequence $S_u^c$ is $E_u^c=g_\theta([e_{c_{q_1}},e_{c_{q_2}},...,e_{c_{q_n}}])$. Consider $S_u^q$ and $S_u^{c}$ under the same user as positive samples, and the search history of other users as negative samples. The definition of alignment loss is as follows:

\begin{equation}
\begin{aligned}
\mathcal{L}_{align}=-[log\frac{exp(E_u^q, E_u^c)/\tau}{\sum_{u' \in U}exp(E_u^q, E_{u'}^c)/\tau} + log\frac{exp(E_u^c, E_u^q)/\tau}{\sum_{u' \in U}exp(E_u^c, E_{u'}^q)/\tau}]
\end{aligned}
\end{equation}

Next, we will perform semantic mapping on each behavioral representation in the mixed sequence $S_u$ to obtain a semantically sequential representation $\tilde{E}_u$, it should be noted that considering the large amount of side info in search behavior, in order to alleviate the problem of domain information asymmetry in semantic partitioning, the search behavior here only uses the average pooled embedding representation of clicked item sequences. The formal definition of $\tilde{E}_u$ is as follows:

\begin{equation}
\begin{aligned}
\tilde{e}_{x_t} &= \left\{
\begin{array}{ll}
g_\theta(e_{i_t}), & \text{if } b_t=1 \\
g_\theta(Mean(c_{q_t})), & \text{if } b_t=0
\end{array}
\right. 
\end{aligned}
\end{equation}

Afterwards, we will divide the subsequences based on the global semantic cluster $CQ$. Specifically, we will calculate the similarity between each behavior's semantic representation of $\tilde{E}_u$ and each cluster center to obtain a similarity matrix $T_u \in R^{L\times k}$, where $L$ is the sequence length. The semantic label $l_i$ for each position $i$ is the index of the cluster center with the highest similarity. Finally, we divide the subsequence $S_u$ based on each position's label. This process can be formally defined as:

\begin{equation}
\begin{aligned}
T_{ij}&=\tilde{e}_i\hat{c}_{q_j}^T, i \leq |S_u|, j \leq |CQ| \\
l_i &= argmax_jT_{ij} \\
S_u^j &= \{x_i|l_i=j\}
\end{aligned}
\end{equation}

Then, based on the candidate item $e_{i_{N+1}}$, we extract semantic multi interests $\{S_u^j\}_{j=1}^k$ for each semantic subsequence $\{I_{sem}^j\}_{j=1}^k$ through target attention.

\begin{equation}
\begin{aligned}
w_j &= softmax(H_u^{j}W_{j}e_{i_{N+1}}), \quad I_{sem}^j = w_{j}H_u^{j} \\
H_u^j &= [e_{i_1}, e_{i_2}, ..., e_{i_N}], i_j \in S_u^j
\end{aligned}
\end{equation}

Where $W_j$ is the learnable parameter, and $w_j$ is the attention score. After extracting the corresponding semantic interests for each subsequence, we perform interest fusion. Here we use target attention to weight and aggregate multiple semantic interests based on the candidate item $i_{N+1}$, resulting in the final semantic interest $I_{sem}$.

\begin{equation}
\begin{aligned}
w_{sem}^j &= softmax(I_{sem}^jW_{sem}e_{i_{N+1}}), \quad I_{sem} = \sum_{j=1}^kw_{sem}^jI_{sem}^j
\end{aligned}
\end{equation}

Due to the fact that semantic multi-interest is based on semantic clusters, in order to make semantic clustering more reasonable, we have designed orthogonal loss for semantic multi-interest to reduce redundancy.

\begin{equation}
\begin{aligned}
\mathcal{L}_{orth}=\frac{1}{k(k-1)}\sum_{i\neq j}cos^2(I_{sem}^i, I_{sem}^j)
\end{aligned}
\end{equation}

Then, we adaptively fuse the search interest $I_s$, recommendation interest $I_r$, cross interest $I_{cross}$, and semantic interest $I_{sem}$ on the structural side. Here, we still use target attention to weight and aggregate the embedding representations of candidate items $i_{N+1}$ to obtain the final user interest $I_u$.

\begin{equation}
\begin{aligned}
w_{inter} &= softmax(I_{inter}W_{inter}e_{i_{N+1}}), \quad I_{u} = w_{inter}I_{inter} \\
I_{inter} &= Concat([I_s, I_r, I_{cross}, I_{sem}])
\end{aligned}
\end{equation}

Where $W_{inter} \in R^{4d \times d}$ is a learnable parameter.

\subsection{Multi-Task Prediction}
Considering the difference in data distribution between search and recommendation tasks, we adopt a multi-task prediction approach to separately predict search and recommendation tasks. Specifically, we introduce a progressive layered extraction (PLE) model~\cite{tang2020progressive} to achieve S\&R task prediction by setting up shared experts and task experts. We have designed a learnable query $q_\phi \in R^d$ for recommendation tasks. In addition, we extract mixed sequence representation $e_{seq}$ through mask pooling as side info. The input $X_b=concat([e_u,e_{i_{N+1}},e_q,I_{u},e_{seq}])$ of the multi-task prediction model is mainly composed of user representation $e_u$, candidate item representation $e_{i_{N+1}}$, query representation $e_q$ (using $q_\phi$ instead for recommendation tasks), user interest $I_{u}$, and mixed sequence representation $e_{seq}$.

We set the number of search experts $O_s$, recommendation experts $O_r$, and shared experts $O_c$ as $N_s$, $N_r$, and $N_c$, respectively. The output of the prediction layer contains two scores, one is the predicted score for the search task, and the other is the predicted score for the recommendation task. The calculation method is as follows:

\begin{equation}
\begin{aligned}
\hat{y}_{u,i,q}^s = \sum_{o \in O_s}g_o^s(X_b)o(X_b) + \sum_{o \in O_c}g_o^s(X_b)o(X_b)  \\
\hat{y}_{u,i}^r = \sum_{o \in O_r}g_o^r(X_b)o(X_b) + \sum_{o \in O_c}g_o^r(X_b)o(X_b)  \\
\end{aligned}
\end{equation}

The gate networks for two tasks are $g^s(X_b)=softmax(W^sX_b)$ and $g^r(X_b)=softmax(W^rX_b)$, respectively. $W^s \in R^{(N_s+N_c)\times d}$ and $W^r \in R^{(N_r+N_c)\times d}$ are learnable linear layers, where $o(\cdot)$ represents the expert network, which we implement through a multi-layer perceptron (MLP).

\subsection{Training}
We use binary cross-entropy loss as the main task of optimizing the model.

\begin{equation}
\begin{aligned}
\mathcal{L}_{bce}^s&=-\frac{1}{|D_s|}\sum_{(u,i,q) \in D_s}y_{u,i,q}^slog(\hat{y}_{u,i,q}^s)+(1-y_{u,i,q}^s)log(1-\hat{y}_{u,i,q}^s) \\
\mathcal{L}_{bce}^r&=-\frac{1}{|D_r|}\sum_{(u,i) \in D_r}y_{u,i}^rlog(\hat{y}_{u,i}^r)+(1-y_{u,i}^r)log(1-\hat{y}_{u,i}^r)
\end{aligned}
\end{equation}

Where $\mathcal{L}_{bce}^s$ and $\mathcal{L}_{bce}^r$ represent the BCE loss for search and recommendation tasks, respectively. The total loss of the model is as follows:

\begin{equation}
\begin{aligned}
\mathcal{L}_s &= \mathcal{L}_{bce}^s + \alpha\mathcal{L}_{cl} + \beta\mathcal{L}_{orth} + \gamma\mathcal{L}_{align} \\
\mathcal{L}_r &= \mathcal{L}_{bce}^r + \alpha\mathcal{L}_{cl} + \beta\mathcal{L}_{orth} + \gamma\mathcal{L}_{align} \\
\mathcal{L} &= \mathcal{L}_r + \mu\mathcal{L}_s + \eta||\Theta||_2
\end{aligned}
\end{equation}

Where $\mu$ is used to balance two tasks, and $\eta$ is L2 regularization.

\section{Experiment}
We conducted a series of experiments on two open source datasets to evaluate our model. And answered the following questions: 
\begin{itemize}[leftmargin=*] 
    \item \textbf{RQ1}: How effective is our method compared to other baseline models;
    \item \textbf{RQ2}: The impact of the main modules of our model on overall performance;
    \item \textbf{RQ3}: The impact of different values of key hyperparameters on model performance.
\end{itemize}

\subsection{Experiment Setting}
\subsubsection{Datasets}
We conducted experiments using two open-source datasets, Amazon and KuaiSAR, and the statistical data of the datasets are shown in Table 1.

\textbf{KuaiSAR~\cite{sun2023kuaisar}:}\footnote{https://kuaisar.github.io/} A real-world dataset collected from Kuaishou, a popular short-video platform in China. It contains authentic user S\&R behaviors, including explicit search queries and implicit recommendation clicks. We follow the preprocessing protocol in ~\cite{shi2024unisar, si2023search}: filter users/items with fewer than 5 interactions, and split the data into training (first 7 days), validation (1 day), and test (last 2 days) sets. 

\textbf{Amazon Kindle Store~\cite{he2016ups, mcauley2015image}:}\footnote{http://jmcauley.ucsd.edu/data/amazon/} A widely used semi-synthetic cross-scenario dataset derived from the Amazon review dataset. The search behaviors are generated by extracting keywords from product metadata and user reviews and splitting the data using a leave-one-out strategy: the most recent interaction for testing, the second most recent for validation, and the rest for training.


\begin{table}[t]
\centering
\caption{Statistics of the datasets}
\label{tab:dataset}
\begin{tabular}{lccccc}
\toprule
Dataset & \#Users & \#Items & \#Queries & \#Action-S & \#Action-R \\
\midrule
KuaiSAR & 25,877 & 6,890,707 & 453,667 & 5,059,169 & 14,605,716 \\
Amazon  & 68,223 & 61,934    & 4,298   & 934,664   & 989,618 \\
\bottomrule
\end{tabular}
\end{table}



\definecolor{mijsrbg}{gray}{0.92}

\begin{table*}[t]
\centering
\caption{Comparisons of the overall \textit{recommendation} performance of different methods on both datasets. The best and the second-best methods are highlighted in bold and underlined fonts, respectively. \textbf{UniSAR(paper)} refers to the metrics reported in the original paper; \textbf{UniSAR(our)} refers to the results reproduced by our team. (\textit{t}-test, \textit{p}-value< 0.01)}
\label{tab:rec-performance}
\small
\setlength{\tabcolsep}{4pt}

\begin{tabular}{c|c|ccccc|ccccc|c}
\toprule
\multirow{2}{*}{\textbf{Datasets}} &
\multirow{2}{*}{\textbf{Metric}} &
\multicolumn{5}{c|}{\textbf{Sequential Recommendation}} &
\multicolumn{6}{c}{\textbf{Joint Search \& Recommendation}} \\
\cmidrule(lr){3-7}\cmidrule(lr){8-13}
& &
DIN & GRU4Rec & SASRec & BERT4Rec & FMLP-Rec &
JSR & USER & UnifiedSSR & UniSAR (paper) & UniSAR (our) &
\cellcolor{gray!15} \textbf{MIJSR} \\
\midrule
\multirow{5}{*}{\textbf{KuaiSAR}}
 & HR@1
 & 0.1629 & 0.1097 & 0.1249 & 0.1061 & 0.1370
 & 0.1754 & 0.1489 & 0.1225 & \textbf{0.1990} & 0.1853 & \cellcolor{gray!15} \underline{0.1972} \\
 & HR@5
 & 0.4509 & 0.3764 & 0.4065 & 0.3699 & 0.4292
 & 0.4791 & 0.4086 & 0.3981 & \textbf{0.5169} & 0.4887 & \cellcolor{gray!15} \underline{0.5115} \\
 & HR@10
 & 0.6179 & 0.5788 & 0.6007 & 0.5885 & 0.6159
 & 0.6453 & 0.5627 & 0.5939 & \textbf{0.6792} & 0.6547 & \cellcolor{gray!15} \underline{0.6746} \\
 & NDCG@5
 & 0.3104 & 0.2435 & 0.2671 & 0.2381 & 0.2851
 & 0.3315 & 0.2820 & 0.2617 & \textbf{0.3632} & 0.3418 & \cellcolor{gray!15} \underline{0.3593} \\
 & NDCG@10
 & 0.3643 & 0.3087 & 0.3298 & 0.3083 & 0.3453
 & 0.3853 & 0.3318 & 0.3249 & \textbf{0.4158} & 0.3955 & \cellcolor{gray!15} \underline{0.4122} \\
\midrule
\multirow{5}{*}{\textbf{Amazon}}
 & HR@1
 & 0.2159 & 0.1725 & 0.2059 & 0.2481 & 0.1991
 & 0.2346 & 0.2361 & 0.2013 & 0.3010 & \underline{0.3066} & \cellcolor{gray!15} \textbf{0.3172} \\
 & HR@5
 & 0.5170 & 0.4949 & 0.5295 & 0.5311 & 0.5356
 & 0.5467 & 0.5441 & 0.5196 & 0.5874 & \underline{0.5971} & \cellcolor{gray!15} \textbf{0.6132} \\
 & HR@10
 & 0.6525 & 0.6548 & 0.6772 & 0.6658 & 0.6879
 & 0.6779 & 0.6854 & 0.6707 & 0.7020 & \underline{0.7162} & \cellcolor{gray!15} \textbf{0.7338} \\
 & NDCG@5
 & 0.3726 & 0.3388 & 0.3747 & 0.3954 & 0.3739
 & 0.3970 & 0.3964 & 0.3662 & 0.4513 & \underline{0.4592} & \cellcolor{gray!15} \textbf{0.4732} \\
 & NDCG@10
 & 0.4165 & 0.3907 & 0.4225 & 0.4390 & 0.4232
 & 0.4396 & 0.4422 & 0.4151 & 0.4885 & \underline{0.4978} & \cellcolor{gray!15} \textbf{0.5123} \\
\bottomrule
\end{tabular}
\end{table*}

\begin{table*}[t]
\centering
\caption{Comparisons of the overall \textit{search} performance of different methods on both datasets. The best and the second-best methods are highlighted in bold and underlined fonts, respectively. \textbf{UniSAR(paper)} refers to the metrics reported in the original paper; \textbf{UniSAR(our)} refers to the results reproduced by our team. (\textit{t}-test, \textit{p}-value< 0.01)}
\label{tab:src-performance}
\resizebox{\textwidth}{!}{%
\begin{tabular}{c|c|cccccc|ccccc|c}
\toprule
 &  & \multicolumn{6}{c|}{\textbf{Personalized Search}} & \multicolumn{6}{c}{\textbf{Joint Search \& Recommendation}} \\
\cmidrule(lr){3-8}\cmidrule(lr){9-14}
\textbf{Datasets} & \textbf{Metric} &
QEM & HEM & AEM & ZAM & TEM & CoPPS &
JSR & USER & UnifiedSSR & UniSAR(paper) & UniSAR(our) & \cellcolor{gray!15} \textbf{MIJSR} \\
\midrule
\multirow{5}{*}{\textbf{KuaiSAR}}
 & HR@1   & 0.2944 & 0.3337 & 0.2703 & 0.2815 & 0.3045 & 0.3117 & 0.4543 & 0.4628 & 0.4389 & \textbf{0.5282} & 0.4535 & \cellcolor{gray!15} \underline{0.5248} \\
 & HR@5   & 0.6020 & 0.6505 & 0.5956 & 0.6117 & 0.6502 & 0.6616 & 0.7162 & 0.7304 & 0.7377 & \underline{0.7476} & 0.7412 & \cellcolor{gray!15} \textbf{0.7811} \\
 & HR@10  & 0.7182 & 0.7653 & 0.7182 & 0.7344 & 0.7632 & 0.7707 & 0.7961 & 0.8149 & 0.8320 & 0.8369 & \underline{0.8405} & \cellcolor{gray!15} \textbf{0.8632} \\
 & NDCG@5 & 0.4575 & 0.5029 & 0.4415 & 0.4560 & 0.4887 & 0.4977 & 0.5962 & 0.6069 & 0.5991 & \underline{0.6417} & 0.6073 & \cellcolor{gray!15} \textbf{0.6635} \\
 & NDCG@10& 0.4953 & 0.5400 & 0.4812 & 0.4959 & 0.5254 & 0.5331 & 0.6221 & 0.6342 & 0.6297 & \underline{0.6708} & 0.6397 & \cellcolor{gray!15} \textbf{0.6902} \\
\midrule
\multirow{5}{*}{\textbf{Amazon}}
 & HR@1   & 0.2772 & 0.2497 & 0.2916 & 0.2954 & 0.4090 & 0.4052 & 0.3176 & 0.4123 & 0.3663 & \underline{0.5343} & 0.5329 & \cellcolor{gray!15} \textbf{0.5953} \\
 & HR@5   & 0.7100 & 0.6778 & 0.7095 & 0.7109 & 0.8185 & 0.8169 & 0.7038 & 0.7631 & 0.7744 & \underline{0.8190} & 0.8149 & \cellcolor{gray!15} \textbf{0.8265} \\
 & HR@10  & 0.8186 & 0.8267 & 0.8443 & 0.8468 & \textbf{0.9051} & \underline{0.9051} & 0.8225 & 0.8697 & 0.8812 & 0.8977 & 0.8946 & \cellcolor{gray!15} 0.9042 \\
 & NDCG@5 & 0.5066 & 0.4736 & 0.5114 & 0.5147 & 0.6303 & 0.6281 & 0.5173 & 0.6000 & 0.5847 & \underline{0.6875} & 0.6841 & \cellcolor{gray!15} \textbf{0.7160} \\
 & NDCG@10& 0.5422 & 0.5221 & 0.5554 & 0.5590 & 0.6587 & 0.6570 & 0.5563 & 0.6348 & 0.6196 & \underline{0.7132} & 0.7101 & \cellcolor{gray!15} \textbf{0.7413} \\
\bottomrule
\end{tabular}%
}
\end{table*}

\subsubsection{Baselines}
We compared MIJSR with 15 representative baseline methods and divided them into three categories to highlight its advantages in joint S\&R modeling:

\textbf{Sequential Recommendation Models:} These methods only leverage recommendation behavior data and do not utilize search signals: (1) DIN: Uses a local activation unit to adaptively model user interests ~\cite{zhou2018deep}. (2) GRU4Rec: Applies Gated Recurrent Units (GRUs) to model session-based sequential dependencies ~\cite{hidasi2015session}. (3) SASRec: Adopts Transformer for sequential recommendation ~\cite{kang2018self}. (4) BERT4Rec: Uses a bidirectional Transformer with a cloze objective for sequential modeling ~\cite{sun2019bert4rec}. (5) FMLP-Rec: An all-MLP architecture with frequency-domain filters for sequential recommendation ~\cite{zhou2022filter}.  

\textbf{Personalized Search Models:} These methods focus on search task optimization and do not incorporate recommendation data: (1) QEM: A query-item matching model that only considers semantic relevance ~\cite{ai2019zero}. (2) HEM: A hierarchical embedding model for personalized product search ~\cite{ai2017learning}. (3) AEM: Uses attention to aggregate user historical behaviors for search ~\cite{ai2019zero}. (4) ZAM: Enhances AEM with a zero vector to control personalization degree ~\cite{ai2019zero}. (5) TEM: Replaces the attention layer in AEM with a Transformer encoder ~\cite{bi2020transformer}. (6) CoPPS: Applies contrastive learning to optimize user sequence representation for search ~\cite{dai2023contrastive}. 

\textbf{Joint S\&R Models:} These methods jointly model S\&R behaviors to enhance both tasks: (1) JSR: Optimizes search and recommendation models with a joint loss function ~\cite{zamani2020learning}. (2) USER: Integrates S\&R behaviors into a single sequence and encodes it with a Transformer ~\cite{yao2021user}. (3) UnifiedSSR: Proposes a dual-branch network for cross-scenario and cross-view modeling ~\cite{xie2024unifiedssr}. (4) UniSAR: Effectively models the different types of fine-grained behavior transitions for providing users a unified search and recommendation service. ~\cite{shi2024unisar}. 



\subsubsection{Evaluation Metrics}
We use Hit Ratio (HR) and Normalized Discounted Cumulative Gain (NDCG) as evaluation metrics. We report HR@1, HR@5, HR@10 and NDCG@5, NDCG@10. In the candidate settings of the test set, we pair the ground-truth item with 99 randomly sampled negative items that the user has not interacted with.

\subsubsection{Implementation Details}
We set the embedding dimensions for both KuaiSAR and Amazon datasets to 64, the maximum length for recommendation and search sequences to 30, and the maximum length for query segmentation to 50. The numbers of shared and specific experts, $N_c$, $N_s$, and $N_r$, are set to 4. Select the number of semantic center clusters $k$ from \{5, 10, 15, 20\}. Regarding the loss weights $\alpha$, $\beta$, and $\gamma$ in section 4.4, they are all selected from \{1e-1, 1e-2, 1e-3, 1e-4, 1e-5\}. The weight $\mu$ of the search task is selected from \{0.01, 0.05, 0.1, 0.5, 1.0\}. Set the batch size to 1024, we train all models with 100 epochs and use an early stopping mechanism to avoid overfitting. Adam is used for optimization.

\subsection{Model Performance (RQ1)}
To demonstrate the superiority of our model in search and recommendation tasks, we compared it with many baseline models. As shown in Table ~\ref{tab:rec-performance} and Table ~\ref{tab:src-performance}, our model MIJSR demonstrated excellent performance in both scenarios. In terms of search tasks, MIJSR showed significant improvements in performance compared to the SOTA model UniSAR on both datasets. On the KuaiSAR dataset, HR@5 increased by 4.48\%, NDCG@5 improved by 3.40\%; on the Amazon dataset, HR@5 and NDCG@5 increased by 0.92\% and 4.14\%, respectively. In terms of recommendation tasks, there is a significant performance gain on the Amazon dataset, HR@5 and NDCG@5 improved by 4.39\% and 4.85\%, respectively, while on the KuaiSAR dataset, MIJSR's various indicators were basically the same as the results reported in the UniSAR paper, which is probably due to the multitasking seesaw. However, from the overall effect of the two tasks, MIJSR is better than UniSAR. The comparison with the baseline model fully demonstrates the effectiveness of MIJSR in multi-interest mining from both structural and semantic perspectives.

It is worth noting that the state-of-the-art model UniSAR we compared has a certain deviation from the conclusions of the original paper in terms of local validation of its performance. We believe this is due to the problem of the multitasking seesaw, and the model will make an early stop judgment based on the average value of recommendation tasks' and search tasks' NDCG@5, so it's difficult to fully reproduce the reported results of the original paper. One of the major discrepancies is the evaluation results of the KuaiSAR dataset in recommendation tasks. We investigated the citation status of this paper and found that the DHIM~\cite{zhu2026dual} work recorded similar results for UniSAR to our own work. Therefore, we believe that this reproduction bias should be reasonable, and we will also provide relevant logs in the open source code to prove the validity of this result.

\subsection{Ablation Study (RQ2)}
In order to comprehensively explore the impact of each module of MIJSR on the overall performance of the model, we conducted the following ablation experiments:

\begin{itemize}[leftmargin=*] 
    \item \textbf{w/o $\mathcal{L}_{cl}$}: Delete the contrastive learning task between query and item, explore the impact of the cross-domain behavior fusion module on model performance.
    \item \textbf{w/o $\mathcal{L}_{align}$}: Remove the sequence-level query-item alignment loss and explore the effectiveness of the semantic mapper $g_\theta$ in the absence of guidance.
    \item \textbf{w/o $\mathcal{L}_{orth}$}: Remove the orthogonal loss of semantic multi-interest and explore the effectiveness of semantic clustering in unsupervised situations.
    \item \textbf{w/o structural interests}: Remove all structural interests (search interests, recommendation interests, and cross-interests).
    \item \textbf{w/o semantic interests}: Remove multi semantic interests and explore the impact of semantic interests on model performance.
    \item \textbf{w/o cross interest}: Removing cross interests and exploring the importance of interest mining between cross-domain behaviors.
    \item \textbf{w/o PLE}: Change PLE network to a simple MLP to explore the importance of multi-expert networks for multi-task prediction.
\end{itemize}


\begin{figure}[t]
  \centering
  \begin{subfigure}[t]{0.48\linewidth}
    \centering
    \includegraphics[width=\linewidth]{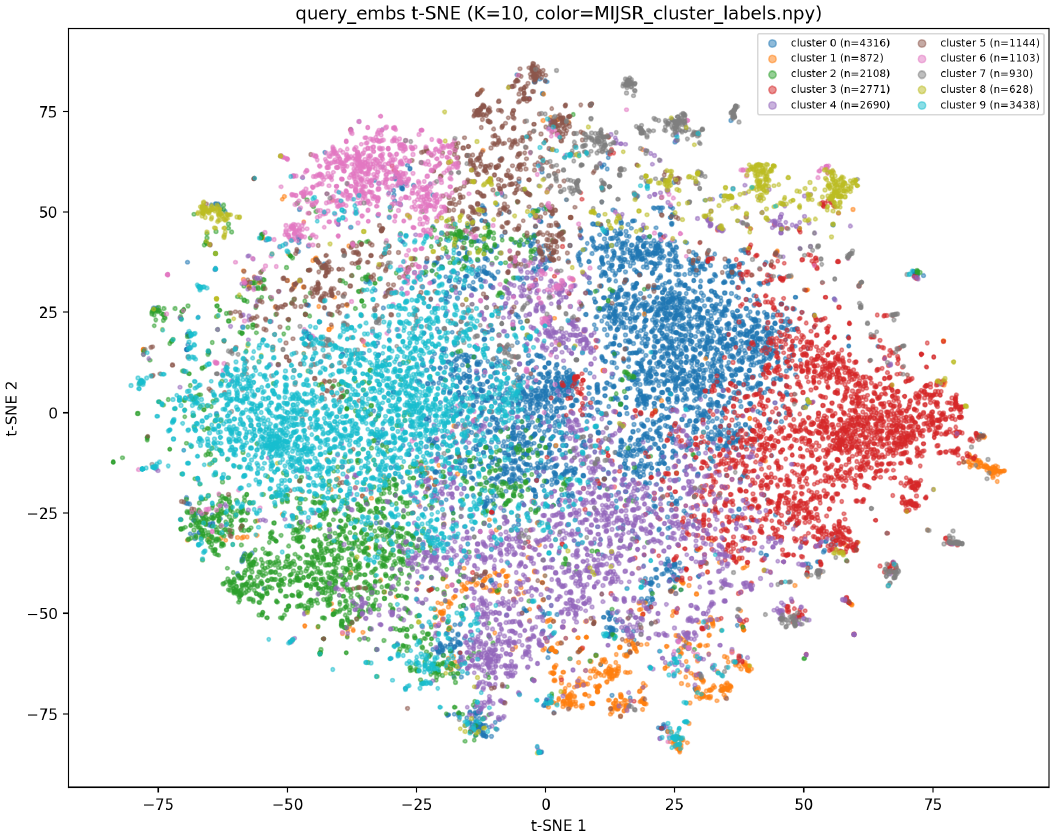}
    \caption{MIJSR}
    \label{fig:a}
  \end{subfigure}
  \hfill
  \begin{subfigure}[t]{0.48\linewidth}
    \centering
    \includegraphics[width=\linewidth]{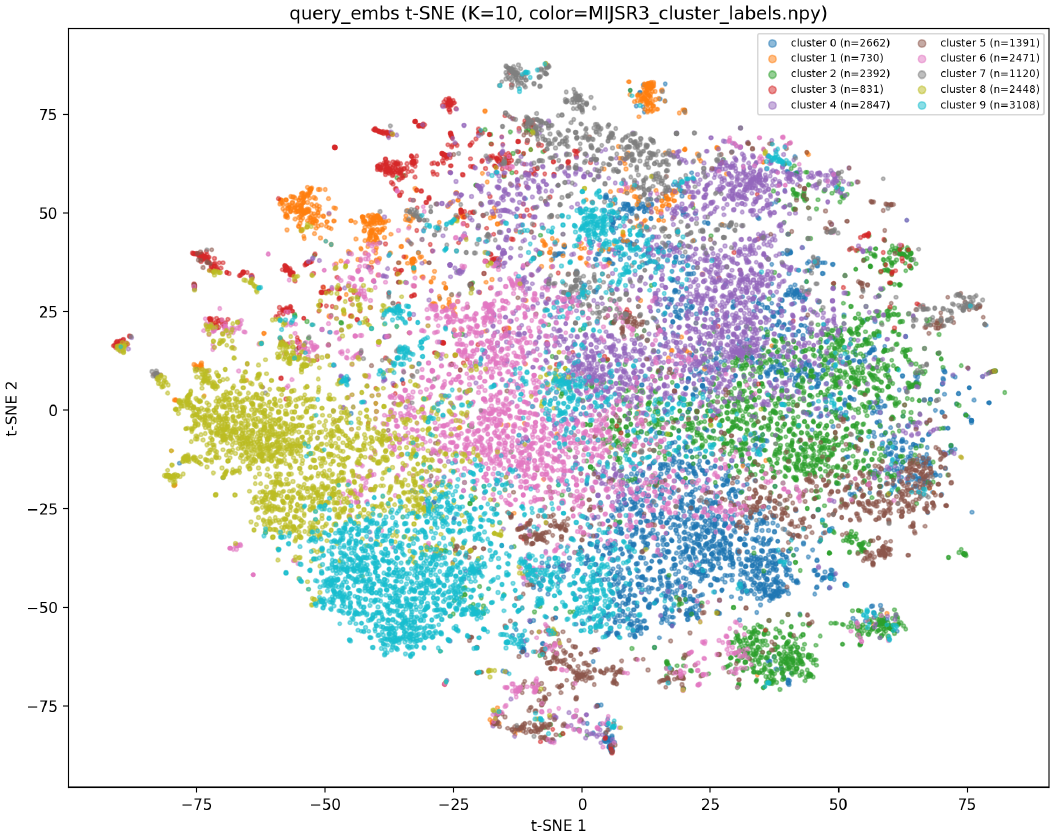}
    \caption{w/o $\mathcal{L}_{orth}$}
    \label{fig:b}
  \end{subfigure}
  \caption{t-SNE visualizations of query embeddings with and without $\mathcal{L}_{orth}$}
  \label{fig:semantic}
  \Description{Comparison of t-SNE visualizations of query embeddings with and without $\mathcal{L}_{orth}$}
\end{figure}

\begin{table*}[t]
\centering
\caption{Ablation study on KuaiSAR and Amazon datasets. ``w/o'' indicates that the corresponding module in MIJSR is removed.}
\label{tab:ablation}
\renewcommand{\arraystretch}{1.08}
\setlength{\tabcolsep}{3.2pt}
\begin{adjustbox}{max width=\textwidth}
\begin{tabular}{l|cc|cc!{\vrule width 0.5pt}cc|cc}
\hline
 & \multicolumn{4}{c!{\vrule width 0.5pt}}{\textbf{KuaiSAR}}
 & \multicolumn{4}{c}{\textbf{Amazon}} \\
\cline{2-5}\cline{6-9}
\textbf{Model}
 & \multicolumn{2}{c|}{Recommendation} & \multicolumn{2}{c|}{Search}
 & \multicolumn{2}{c|}{Recommendation} & \multicolumn{2}{c}{Search} \\
\cline{2-3}\cline{4-5}\cline{6-7}\cline{8-9}
 & HR@5 & NDCG@5 & HR@5 & NDCG@5
 & HR@5 & NDCG@5 & HR@5 & NDCG@5 \\
\hline
\textbf{MIJSR}
 & \textbf{0.5115} & \textbf{0.3593} & \textbf{0.7811} & \textbf{0.6635}
 & \textbf{0.6132} & \textbf{0.4732} & \textbf{0.8265} & \textbf{0.7160} \\
\hdashline
w/o $\mathcal{L}_{cl}$
 & 0.4970 & 0.3442 & 0.7703 & 0.6474
 & 0.5591 & 0.4179 & 0.6301 & 0.4792 \\
w/o $\mathcal{L}_{align}$
 & 0.4618 & 0.3346 & 0.7537 & 0.6180
 & 0.5959 & 0.4585 & 0.8235 & 0.6867 \\
w/o $\mathcal{L}_{orth}$
 & 0.5135 & 0.3596 & 0.7774 & 0.6539
 & 0.5979 & 0.4574 & 0.8200 & 0.6868 \\
\hdashline
w/o structural interests
 & 0.4700 & 0.3267 & 0.7542 & 0.6586
 & 0.5904 & 0.4559 & 0.8071 & 0.6760 \\
w/o semantic interests
 & 0.4013 & 0.3506 & 0.7593 & 0.6428
 & 0.5758 & 0.4486 & 0.8093 & 0.6892 \\
w/o cross interest
 & 0.5090 & 0.3588 & 0.7712 & 0.6594
 & 0.5999 & 0.4612 & 0.8203 & 0.6877 \\
\hdashline
w/o PLE
 & 0.4937 & 0.3474 & 0.7744 & 0.6463
 & 0.5882 & 0.4491 & 0.8137 & 0.6811 \\
\hline
\end{tabular}
\end{adjustbox}
\end{table*}

\begin{figure*}[t]
    \centering
    \includegraphics[width=1\linewidth]{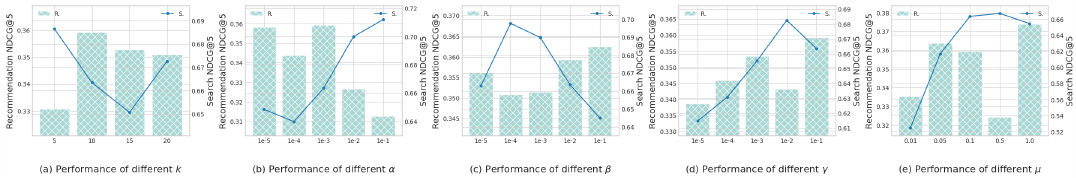}
    \caption{Performance comparison of different hyperparameters under various values.}
    \label{hyper}
\end{figure*}

The experimental results are shown in Table ~\ref{tab:ablation}, from which we can observe the following conclusions:

\begin{itemize}[leftmargin=*] 
\item The performance of the model has decreased after deleting $\mathcal{L}_{cl}$, and the trend of decline is very significant on the Amazon dataset, indicating the necessity of aligning queries and items to enhance behavior representation before conducting mixed sequence modeling.
    \item The performance of the model also decreased after deleting $\mathcal{L}_{align}$, but it was not significant in the search task. This is because the alignment task is mainly used to map the ID representation of the item to the query semantic space, while the item representation and query representation on the search side are represented through $\mathcal {L}_{cl}$ which can already align well, but it will have a significant impact on the items on the recommendation side.
    \item Delete $\mathcal{L}_{orth}$, the performance of the model slightly decreased. This task promotes more reasonable semantic clustering by making semantic multi interests as orthogonal as possible. Here, we also performed query embedding sampling on the KuaiSAR dataset, randomly extracting 10000 queries from all queries. Based on the checkpoint of the model, we extracted the clustering center and query embedding and then visualized the clustering effect through t-SNE. As shown in Fig. ~\ref{fig:semantic}, after removing $\mathcal{L}_{orth}$, there is severe center aliasing and most clusters are loose.
    \item Removing multi-interest modeling on the structural side significantly reduces model performance, indicating that relying solely on semantics to divide multi interests is not enough. It is also necessary to explore users' potential preferences from the perspective of behavior types themselves.
    \item Removing semantic multi interests significantly reduces model performance, which further illustrates the importance of query in understanding user preferences and the necessity of semantic segmentation of user multi-interests.
    \item Removing cross interests results in a slight decrease in model performance. This indicates that the transformation of domain behavior can indirectly reflect user interests.
    \item Replacing the PLE layer for multi-task prediction with a simple MLP network will significantly reduce model performance, as PLE can alleviate multi-task negative transfer during joint training while retaining cross-task knowledge.
\end{itemize}


\subsection{Hyperparameter Study (RQ3)}
In this section, we explore the sensitivity of MIJSR to some key hyperparameters on the KuaiSAR dataset. The relevant experimental results are shown in Fig. ~\ref{hyper}.

\subsubsection{Effects of the number of semantic clusters $k$}

As shown in Fig. ~\ref{hyper}(a), when $k$ increases from 5 to 10, the recommendation performance improves significantly, while the search performance exhibits a slight decline. However, as $k$ further increases to 15 and 20, the performance of both tasks deteriorates. These results indicate that the granularity of semantic interest partitioning plays a crucial role in model performance. When $k$ is relatively small, distinct semantic interests are likely to be grouped into the same cluster, resulting in overly coarse semantic interest representations. In contrast, when $k$ becomes too large, some interest clusters contain only a few behaviors, leading to sparse and unstable semantic interest representations. Moreover, the increased number of clusters introduces additional noise during the interest fusion stage, which negatively affects both search and recommendation performance. Overall, the model achieves the best trade-off between the two tasks when $k=10$.

\subsubsection{Effects of $\mathcal{L}_{cl}$ loss weight $\alpha$}

As shown in Fig. ~\ref{hyper}(b), as $\alpha$ increases from 1e-5 to 1e-3, the recommendation performance consistently improves, while the search performance also exhibits an overall upward trend. However, when $\alpha$ is further increased, the performance of both tasks begins to decline. This observation demonstrates that query-item contrastive learning plays a positive role in unifying the representations of search and recommendation behaviors. A moderate increase in $\alpha$ enhances the semantic consistency between queries and clicked items, enabling search and recommendation behaviors to be mapped into a more unified representation space, which benefits mixed-sequence modeling. Nevertheless, when $\alpha$ becomes excessively large, the contrastive learning objective dominates the optimization process, causing the model to focus excessively on representation alignment while neglecting the primary prediction objectives. As a result, both recommendation and search performance deteriorate.

\subsubsection{Effects of $\mathcal{L}_{orth}$ loss weight $\beta$}

As shown in Fig. ~\ref{hyper}(c), when $\beta$ increases from 1e-5 to 1e-4, the search performance improves substantially, whereas further increases in $\beta$ lead to performance degradation. In contrast, the recommendation performance remains relatively stable across different settings. The objective of $\mathcal{L}_{orth}$ is to reduce redundancy among different semantic interests, encouraging each interest cluster to learn a more independent representation. When $\beta$ is too small, the overlap among interests cannot be effectively constrained, causing multiple interest clusters to capture similar patterns. Conversely, an excessively large $\beta$ overemphasizes the distinction between interests and disrupts the inherent correlations that should exist among them, resulting in overly fragmented interest representations and degraded performance.

\subsubsection{Effects of $\mathcal{L}_{align}$ loss weight $\gamma$}

As shown in Fig. ~\ref{hyper}(d), as $\gamma$ increases from 1e-5 to 1e-2, the performance of both tasks improves significantly. However, further increasing $\gamma$ to 1e-1 leads to a decline in performance. These results suggest that an appropriate alignment constraint helps the semantic mapping network $g_\theta$ effectively project item representations into the query semantic space, thereby improving the accuracy of semantic clustering and semantic interest partitioning. However, when $\gamma$ becomes excessively large, the model over-optimizes the semantic alignment objective at the expense of the discriminative capability required for recommendation and search tasks, ultimately resulting in inferior overall performance.

\subsubsection{Effects of search task loss weight $\mu$}

As shown in Fig. ~\ref{hyper}(e), as $\mu$ increases from 0.01 to 0.5, the search performance consistently improves, while the recommendation performance first improves and then declines. When $\mu$ is further increased to 1.0, the search performance drops slightly, whereas the recommendation performance recovers significantly. This phenomenon reflects the trade-off inherent in multi-task learning between search and recommendation objectives. When $\mu$ is small, model optimization is dominated by the recommendation task, preventing the search task from receiving sufficient training signals and resulting in relatively poor search performance. As $\mu$ increases, the search task receives more gradient information, leading to gradual performance improvements. Meanwhile, the explicit interest signals provided by search behaviors can also benefit recommendation performance through shared representation learning. However, when $\mu$ becomes excessively large, the model focuses disproportionately on the search objective, suppressing the recommendation task and causing performance degradation. Therefore, the experimental results indicate that a moderate search task weight can better balance the two objectives and maximize the benefits of joint modeling.











\section{Conclusion}


Existing joint search-recommendation models often overlook diverse interests in mixed behavior sequences. We propose MIJSR, which models and adaptively integrates user interests from structural and semantic perspectives. Structurally, it captures search, recommendation, and cross-domain interests to model intra- and inter-domain transitions. Semantically, it leverages queries as explicit preference signals, clusters the query space, segments mixed sequences into semantic subsequences, and extracts corresponding interests. Extensive experiments demonstrate its effectiveness.


\begin{acks}
This work was supported in part by the National Natural Science Foundation of China under Grant No. 62676163 and 62276110.
\end{acks}

\section{Ethical Considerations}
Our work uses publicly available anonymized datasets for research purposes only. While the proposed joint search and recommendation framework improves personalized modeling, it may also increase risks related to user profiling, behavioral bias amplification, and filter bubbles. In addition, such techniques could potentially be misused for manipulative personalization or targeted advertising. To reduce these risks, our method does not use sensitive personal attributes, and we encourage future real-world deployments to adopt privacy-preserving mechanisms, fairness-aware modeling, and user transparency controls.

\bibliographystyle{ACM-Reference-Format}
\bibliography{reference}










\end{document}